\documentclass[aps, prl, twocolumn, superscriptaddress]{revtex4-1}
\usepackage{epsfig}
\usepackage{amsmath}
\usepackage{amssymb}
\usepackage{subfigure}
\usepackage{bm}   
\usepackage{verbatim} 
\usepackage{natbib}

\begin{document}

\title{A geometric solution to the coincidence problem,\\ and the size of the landscape as the origin of hierarchy}
\date{November 3, 2010}
\author{Raphael Bousso}
\affiliation{Center for Theoretical Physics and Department of Physics\\
\ \  University of California, Berkeley, CA 94720-7300, U.S.A.}
\affiliation{Lawrence Berkeley National Laboratory, Berkeley, CA 94720-8162,
  U.S.A.}
\author{Ben Freivogel}
\affiliation{Center for Theoretical Physics and Laboratory for Nuclear Science\\
\ \ Massachusetts Institute of Technology, Cambridge, MA 02139, U.S.A.}
\author{Stefan Leichenauer}
\author{Vladimir Rosenhaus}
\affiliation{Center for Theoretical Physics and Department of Physics\\
\ \  University of California, Berkeley, CA 94720-7300, U.S.A.}
\affiliation{Lawrence Berkeley National Laboratory, Berkeley, CA 94720-8162,
  U.S.A.}
\pacs{}
\keywords{}
\begin{abstract}
Without assuming necessary conditions for observers such as galaxies or entropy production, we show that the causal patch measure predicts the coincidence of vacuum energy and present matter density.  Their common scale, and thus the enormous size of the visible universe, has its origin in the number of metastable vacua in the landscape.
\end{abstract}
\maketitle
\paragraph{Introduction}

The smallness of the cosmological constant, $\Lambda$, has long posed a hierarchy problem: why is the energy of the vacuum at least sixty orders of magnitude smaller than the natural value expected from the Standard Model~\cite{Wei89,Pol06,Bou07}? 

The actual magnitude of the cosmological constant was measured more recently~\cite{SN1,SN2,WMAP5}; it poses, in addition, a coincidence problem.  The vacuum energy,
\begin{equation}
\rho_\Lambda \equiv \frac{\Lambda}{8\pi}\approx  1.4\times 10^{-123}~,
\label{eq-s}
\end{equation}
is comparable in magnitude to the present matter density. (We use Planck units, with $c=\hbar=G=1$.)  This constitutes a coincidence of two {\em a priori\/} unrelated timescales:
\begin{equation}
t_\Lambda \sim t_{\rm obs}~,
\label{eq-c}
\end{equation}
where $t_\Lambda\sim \rho_\Lambda ^{-1/2}$ is the time at which vacuum energy begins to dominate, and $t_{\rm obs}$ is the time at which observers exist.

In a theory that contains a sufficiently dense discretuum of values of $\Lambda$, anthropic arguments can explain the smallness of $\Lambda$. Weinberg~\cite{Wei87} argued that observers require galaxies, which form only if $t_\Lambda \gtrsim t_{\rm gal}$, where $t_{\rm gal}$ is the time at which density perturbations grow nonlinear.  This approach successfully predicted a nonvanishing cosmological constant, and it explains the coincidence that $t_\Lambda\sim t_{\rm gal}$.  It has the following limitations:
\begin{itemize} 
\item It does not explain the coincidence $t_{\rm obs} \sim t_\Lambda$.  (There is no reason {\em a priori\/} why $t_{\rm obs}$ should not vastly exceed $t_{\rm gal}$.)
\item It applies only to observers whose existence depends on the formation of galaxies.
\item It actually favors values of $\Lambda$ a few orders of magnitude larger than the observed value.
\item It explains one unnaturally large timescale, $t_\Lambda$, in terms of another, $t_{\rm gal}$.  This scale, in turn, arises from a combination of unnaturally small quantities whose fundamental origin is not clear (the strength of primordial density perturbations and the temperature at matter-radiation equality).  The correlation of $t_\Lambda$ with $t_{\rm gal}$ is a highly nontrivial prediction, but it does not address the fundamental origin of their enormous scale.
\item The prediction was based on a cosmological measure, observers per baryon, that has been ruled out phenomenologically~\cite{BouFre06b}.
\end{itemize} 
The causal patch measure transcends the above limitations.  It explains the coincidence (\ref{eq-c}) from geometric properties, assuming only that observers are made of matter or radiation; and it relates the absolute scale, (\ref{eq-s}), to the size of the landscape.  

\paragraph{Results} 

Without assuming specific conditions necessary for observers, we will derive three results.  The first, $t_\Lambda \sim t_{\rm obs}$, solves the coincidence problem and predicts that {\em all\/} observers are most likely to find themselves at the onset of vacuum domination, independently of their nature and of $t_{\rm obs}$.  Our second result, $t_{\rm c} \sim t_{\rm obs}$, predicts another coincidence: that the timescale associated with spatial curvature is comparable to $t_{\rm obs}$. Our third result, $t_{\rm obs} \sim \bar {\cal N}^{1/2}$, explains the origin of the enormous phenomenological scales governing the visible universe, such as its size and age, in terms of the number of landscape vacua that can be cosmologically produced and which contain observers, $\bar {\cal N}$.  It can be regarded as a prediction for $\bar {\cal N}$.   We stress that $\bar {\cal N}$ will be smaller than ${\cal N}$, the total number of vacua, due to anthropic and cosmological selection effects, neither of which we will consider in this paper. Intriguingly, preliminary analyses~\cite{BP,DenDou04b} suggest $\log({\cal N}^{1/2})\sim O(100)$ for a class of vacua~\cite{KKLT}.  If this class dominates the landscape, then our prediction implies that $\log\bar{\cal N}$ is not very much smaller than $\log {\cal N}$, and that the fundamental origin of hierarchy lies in the number of topological cycles of complex three-manifolds.

\paragraph{Relation to other work} 

Polchinski~\cite{Pol06} first articulated the question of the ultimate origin of the scale of the cosmological constant in the context of the landscape. Refs.~\cite{Bou06,Bou07} anticipated the answer reached in the present paper but not its derivation.  Less general derivations have been proposed for small subsets of the landscape (vacua that differ from ours only in a few parameters)~\cite{BouHal09,BouLei09}, or under the assumption that observers arise in proportion to the entropy produced~\cite{Bou06b,BouHar10}.  A direct antecedent~\cite{BouHar10} of our arguments employed the causal {\em diamond}, a measure that is somewhat less well-defined than the causal patch, and which suppresses the subtle role played by curvature in the present analysis.

Because the three timescales are well-defined across the entire landscape and no specific anthropic assumptions are made, our results apply to arbitrary observers in arbitrary vacua.   Other parameters like the masses of leptons~\cite{HalNom07} or the timescales of structure formation~\cite{BouHal09}, may be correlated with $t_{\rm obs}$, but they are defined only in small portions of the landscape and will not be considered here.   In a separate publication, we will study other measures, and we will argue that currently no measure is viable in the domain of dependence of a spacelike singularity, such as regions with negative vacuum energy.  Here, we restrict to vacua with $\Lambda>0$.

Our arguments differ significantly from those of Refs.~\cite{BanJoh05,LinVan10}, which appeal to the discreteness of the string landscape to cut off a probability distribution that favors small values of the cosmological constant {\em exponentially\/} in $\Lambda^{-1}$, regardless of $t_{\rm obs}$. These approaches appear to conflict with observation because small $\Lambda$ is selected too strongly.  (Vacua in which observers arise by dynamical evolution from a low-entropy initial state are extremely rare in a realistic landscape,  compared to vacua in which observers can be formed only by quantum fluctuations but with probability greater than $e^{-1/\Lambda}$.  So the vacuum with smallest $\Lambda$ is likely in the latter class, and one predicts observations that are incompatible with a long semiclassical history.)

\paragraph{Derivation}

The relative probability for two outcomes of a cosmological measurement is given by $p_1/p_2=N_1/N_2$, where $N_I$ is the expected number of times each outcome occurs in the universe.  Thus, the $N_I$ play the role of an unnormalized probability distribution.  A distribution $dp/dx$ over a continuous parameter $x$ can be computed as the number $dN$ of outcomes occurring in the range $(x,x+dx)$.

\begin{figure}[tbp]
\centering
   \includegraphics[width=3.5in]{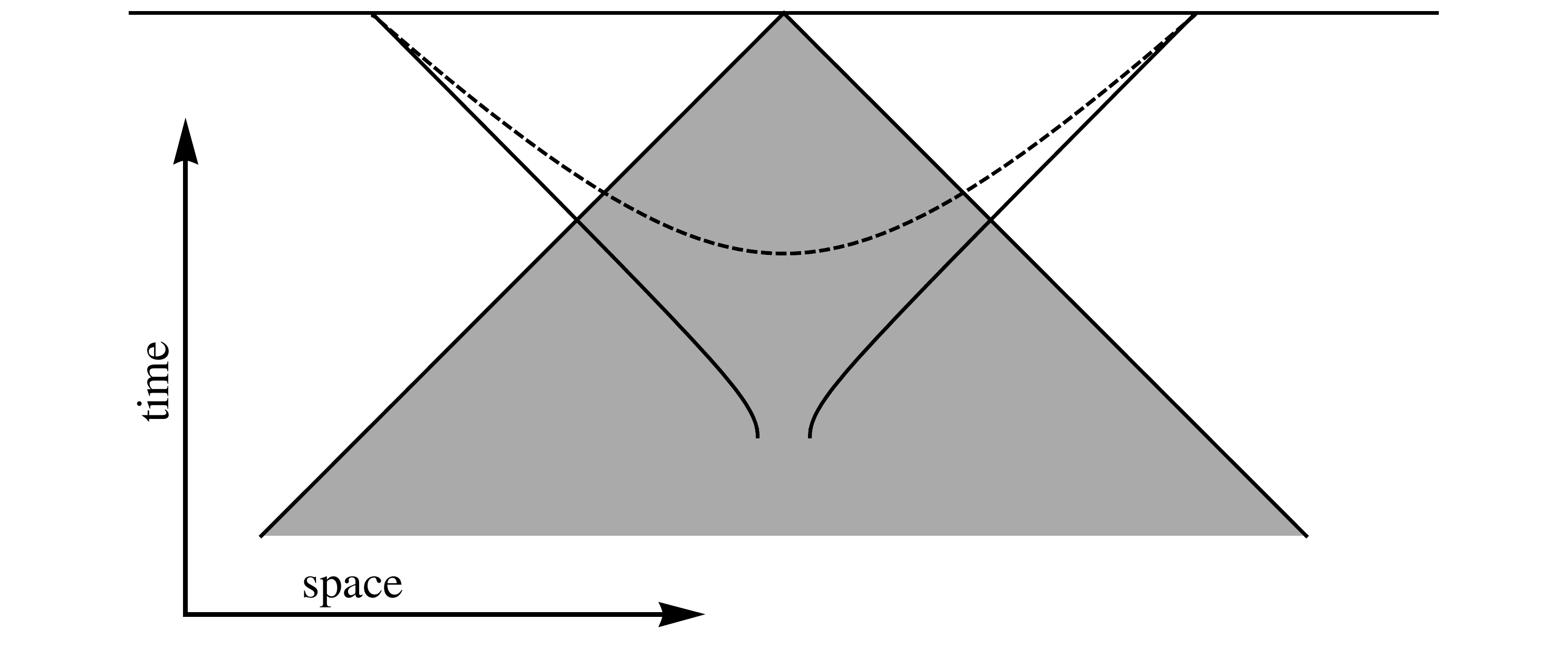}
   \caption{Conformal diagram of a portion of the multiverse.  The dashed line shows an infinite hyperbolic surface of constant FRW time $t_{\rm obs}$.  The causal patch (shaded) restricts attention to the finite portion that lies within the event horizon.}
   \label{fig-cp} 
\end{figure}

The landscape of string theory contains long-lived de~Sitter vacua which give rise to eternal inflation~\cite{BP}. Globally, every experiment and every possible outcome occurs infinitely many times: $N_I=\infty$.  To obtain well-defined relative probabilities, these divergences must be regulated: this is the measure problem of eternal inflation.  Here we consider the causal patch measure~\cite{Bou06,BouFre06a}, which restricts to the causal past of a point on the future boundary of spacetime (see Fig.~\ref{fig-cp}), in which $N_I$ is to be computed.  (Because geometric cutoffs such as the causal patch disrupt the worldlines of some particles, they require a justification in terms of a physical mechanism~\cite{BouFre10b}.  This remains an important open problem which we do not address here.)

Consider an arbitrary observer who lives at an FRW time of order $t_{\rm obs}$.  What order of magnitude for $t_\Lambda$ and $t_{\rm c}$ is he likely to observe?  This is described by the probability distribution over $\log t_\Lambda$ and $\log t_{\rm c}$ at fixed $\log t_{\rm obs}$, which can be written as
\begin{equation}
\frac{dp}{d\log t_\Lambda d\log t_{\rm c}}=\frac{d\tilde p}{d\log t_\Lambda d\log t_{\rm c}} ~n_{\rm obs}(\log t_{\rm c}, \log t_\Lambda; \log t_{\rm obs})~.
\label{eq-split}
\end{equation}
We will begin by computing this distribution; later we will allow $t_{\rm obs}$ to vary as well.

The first factor in Eq.~(\ref{eq-split}) is the ``prior probability''; it corresponds to the expected number of times a vacuum with specified values of $\log t_\Lambda$ and $\log t_{\rm c}$ is nucleated in the causal patch~\cite{BouYan07}.   This is proportional to the number of vacua in the landscape with specified values of $\log t_\Lambda$ and $\log t_{\rm c}$, multiplied by the rate at which such vacua are produced cosmologically from specified initial conditions.  But this rate will be independent of $t_\Lambda$ and $t_{\rm c}$ in the regime of interest ($t_\Lambda \gg 1$).  In the string landscape, a decay changes $\Lambda$ enormously compared to the energy scales associated with $t_\Lambda$ and $t_{\rm c}$ in the daughter vacuum.  Thus, the decay chains leading to vacua of interest have no information about the eventual values of $t_\Lambda$ and $t_{\rm c}$.

The cosmological constant in a metastable vacuum includes large nongravitational contributions, so zero is not a special value for the sum.  Hence, the density of vacua, $d {\cal N} /d\Lambda$, can be Taylor-expanded around $\Lambda=0$.  Vacua with $\Lambda\sim 1$ contain only a few bits of causally connected information~\cite{CEB1}, and hence no complex systems of any kind.  Thus, we may restrict attention to vacua with $\Lambda\ll 1$ and keep only the leading order in the expansion, $d {\cal N} /d\Lambda=$ const~\cite{Wei87}.  With $t_\Lambda\sim \Lambda^{-1/2}$, this implies
\begin{equation}
\frac{d\tilde p}{d\log t_\Lambda d\log t_{\rm c}} = t_\Lambda ^{-2}\, g(\log t_{\rm c})~.
\label{eq-prior}
\end{equation}
Here $g$ encodes the prior probability distribution over the time of curvature domination.

The second factor in Eq.~(\ref{eq-split}) is the number of observers present within the causal patch, at the time $t_{\rm obs}$ in a vacuum with parameters $(t_\Lambda, t_{\rm c})$.  It can be written as
\begin{equation}
n_{\rm obs} = M_{\rm CP}(\log t_\Lambda , \log t_{\rm c} ; \log t_{\rm obs})\, h(\log t_\Lambda , \log t_{\rm c} ; \log t_{\rm obs})~,
\label{eq-mh}
\end{equation}
where $M_{\rm CP}$ is the total matter mass present within the causal patch at the time $t_{\rm obs}$, and $h$ is the number of observers per unit matter mass.  In principle, $h$ is very difficult to compute, but we will argue below that $h$ is trivial in the only regime where it could possibly affect our result.  In vacua with no observers at $t_{\rm obs}$, we set $h=0$.

We now turn to the computation of $M_{\rm CP}$.  Vacua are cosmologically produced as open FRW universes~\cite{CDL} with metric $ds^2 = -dt^2 + a(t)^2 (d\chi^2 +\sinh^2\!\chi\, d \Omega_2^2)$, embedded in an eternally inflating parent vacuum. The boundary of the causal patch is given by the past light-cone from the future end point of a comoving geodesic, which may be chosen to lie at $\chi=0$ by homogeneity, and which coincides with the event horizon for long-lived metastable de~Sitter vacua.  It can be computed as if the de~Sitter vacuum was eternal, as the correction from late-time decay is negligible: $\chi_{\rm CP}(t) = \int_{t}^\infty dt'/a(t')$.

If $t_{\rm c} \lesssim t_\Lambda$, the universe contains a curvature dominated era.   The evolution of the scale factor is governed by the Friedmann equation, $\dot{a}^2/a^2 = t_{\rm c}/a^3 + a^{-2} + t_{\Lambda}^{-2}$, where $t_{\rm c}/a^3\sim \rho_{\rm m}$ is the energy density of pressureless matter.  The remaining terms encode the curvature and the cosmological constant.  (The inclusion of a radiation term would not affect our results qualitatively.)  A piecewise approximate solution is
\begin{equation} \label{eq:a}
a(t)\sim \left\{\begin{array}{ll}
t_{\rm c}^{1/3} t^{2/3}~,& t<t_{\rm c} \\
t~, & t_{\rm c}<t<t_\Lambda \\
t_\Lambda e^{t/t_\Lambda-1}~, & t_\Lambda < t~.
\end{array}\right.
\end{equation} 
Integration yields the comoving radius of the causal patch: 
\begin{equation}
\chi_{\rm CP}(t)\sim \left\{\begin{array}{ll}
1+\log(t_\Lambda/t_{\rm c})+ 3\left[1-(t/t_{\rm c})^{1/3}\right]
~ , \  t<t_{\rm c} \\
1 +\log (t_\Lambda/t)~, \ \ \ \ \ \ \ \ \ \ \ \ \ \ \ \ \ \ \ \ \    t_{\rm c}<t<t_\Lambda \\
e^{-t/t_\Lambda}~, \ \ \ \ \ \  \ \ \ \ \ \ \ \ \ \ \ \ \ \ \ \ \ \ \ \ \ \ \ \ \ \ \ \ \   t_\Lambda<t 
\end{array}\right.
\label{eq-ccp}
\end{equation}
In the regime where curvature never dominates, $t_{\Lambda}\ll t_{\rm c}$, the flat FRW metric is a good approximation: $a(t)\sim t_{\rm c}^{1/3} t^{2/3}~(t<t_\Lambda)$;  $a(t)\sim t_{\rm c}^{1/3} t_\Lambda^{2/3} e^{t/t_\Lambda-1}~(t_\Lambda < t)$. Integration, or setting $t_{\rm c} \to t_\Lambda$ in Eq.~(\ref{eq-ccp}), yields $\chi_{\rm CP}$.

The mass inside the causal patch at the time $t_{\rm obs}$ is $M_{\rm CP} = \rho_{\rm m} a^3 V_{\rm com}= t_{\rm c} V_{\rm com} [\chi_{\rm CP}(t_{\rm obs})]$.  The comoving volume inside a sphere in hyperbolic space, $V_{\rm com}$, can be approximated by $\chi^3$ for $\chi \lesssim 1$ (in the regime $ t_{\Lambda}<t_{\rm obs} $), and by $e^{2\chi}$ for $\chi \gtrsim 1$ (i.e., for $ t_{\Lambda}>t_{\rm obs} $).  Substituting into Eqs.~(\ref{eq-split})--(\ref{eq-mh}), we find the probability distribution 
\begin{widetext}
\begin{equation} 
\frac{d^2p}{d\log t_{\rm c}~d\log t_\Lambda}\sim  
gh\times\left\{\begin{array}{lll}
1/t_{\rm c}~, & t_{\rm obs}<t_{\rm c} < t_\Lambda \ \ & I\\
t_{\rm c} /t_{\rm obs}^2~,& t_{\rm c}<t_{\rm obs}<t_\Lambda  \ \ & II \\
(t_{\rm c} /t_\Lambda ^{2}) e^{-3t_{\rm obs}/t_\Lambda}~, & t_{\rm c}< t_\Lambda< t_{\rm obs} \ \ &III \\
1/t_{\Lambda}~, & t_{\rm obs}<t_{\Lambda} <t_{\rm c}   \ \  & V\\
t_{\Lambda}^{-1} e^{-3t_{\rm obs}/t_\Lambda}~, & t_\Lambda<t_{\rm obs}\mbox{~and~} t_\Lambda < t_{\rm c} \ \ & IV
\end{array}\right.
\label{eq-pcppos}
\end{equation}
\end{widetext}

\begin{figure}[tbp]
\centering
   \includegraphics[width=3in]{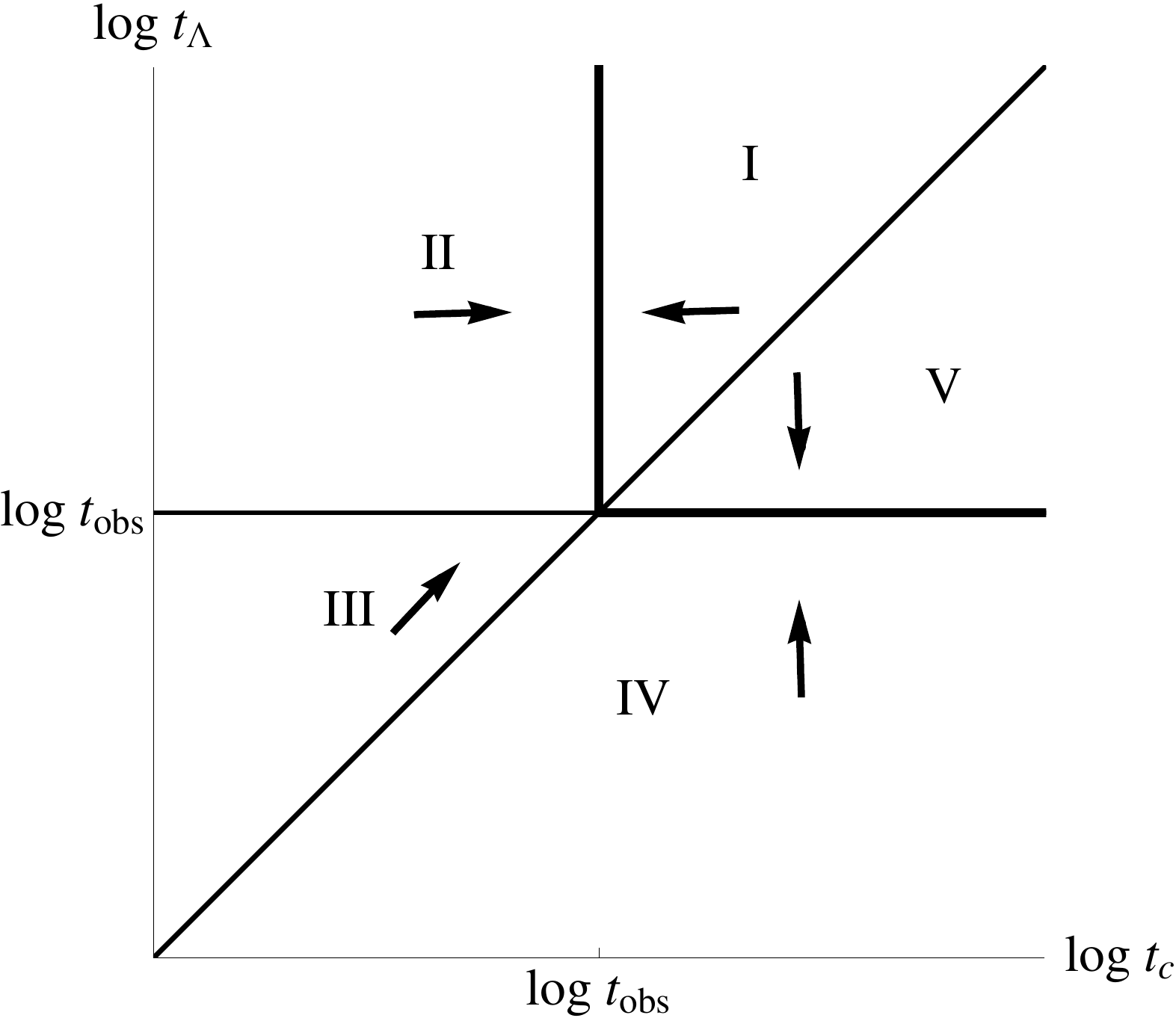}
   \caption{The probability distribution over the timescales of curvature and vacuum domination at fixed observer timescale $t_{\rm obs}$, before the prior distribution over $\log t_{\rm c}$ and the finiteness of the landscape are taken into account.  Arrows indicate directions of increasing probability. The distribution is peaked along two degenerate half-lines (thick).}
   \label{fig-patchflow} 
\end{figure}

For the moment, let us ignore the unknown functions $g$ and $h$, i.e., set $gh$ constant.  Then the probability distribution is a function of powers and exponentials of $t_{\rm c}$ and $t_{\Lambda}$.  Hence, it depends at least exponentially on the variables $\log t_{\rm c}$ and $\log t_{\Lambda}$, so it will be dominated by its maximum.  The maximum can be found by following the gradient flow, which can be thought of as a ``probability force'' ~\cite{HalNom07, BouHal09} pushing $\log t_\Lambda$ and $\log t_{\rm c}$ towards their preferred values.  In Fig.~\ref{fig-patchflow}, arrows indicate whether $\log t_{\rm c}$ and $\log t_\Lambda$ prefer to increase or decrease (or neither).  The arrows do not reflect the precise direction and strength of the gradient.  For example, in region III, the probability grows with $t_{\rm c}$, and since the exponential dominates, it also grows with $t_{\Lambda}$; this is indicated by a diagonal arrow pointing towards larger $\log t_{\rm c}$ and $\log t_\Lambda$. (Recall that we are holding $t_{\rm obs}$ fixed for now.)  By following the arrows, we recognize that the probability density is maximal not at a point, but along two degenerate half-lines of stability: $\log t_{\Lambda} = \log t_{\rm obs},\,\log t_{\rm c}\gtrsim \log t_{\rm obs}$; and $\log t_{\rm c} =\log t_{\rm obs}, \,\log t_\Lambda \gtrsim \log t_{\rm obs}$.

Let us now include the effects of the two prefactors we have so far neglected, beginning with $h$, the number of observers per unit mass.  Crucially, $h$ cannot depend on $t_\Lambda$ for $t_\Lambda\gg t_{\rm obs}$, and it cannot depend on $t_{\rm c}$ for $t_{\rm c} \gg t_{\rm obs}$.  This is because in these regimes, curvature or vacuum energy have no dynamical effect prior to the time $t_{\rm obs}$, so they cannot be measured by anything, including $h$.   For $t_{\rm c} \ll t_{\rm obs}$, we will assume that $h$ is an increasing function of $t_{\rm c}$; similarly, for $t_\Lambda \ll t_{\rm obs} $, we will assume that $h$ is an increasing function of $t_\Lambda$.  This assumption seems extraordinarily weak, in the sense that its converse would be preposterous.  The effects of early curvature domination and vacuum domination are both gravitational, and thus universal.  They limit the amount of matter than can cohere; the smaller $t_{\rm c}$ or $t_\Lambda$, the smaller the largest structure in the universe will be.  To deny our assumption is to claim that on average over many vacua, this disruption is correlated with an {\em increased\/} number of observers per unit mass.  (In fact, an even weaker assumption would suffice, namely that as $t_{\rm c}$ or $t_\Lambda$ are decreased below $t_{\rm obs}$, $h$ does not grow so rapidly as to overwhelm the factors that suppress the probability in this regime.)

We thus arrive at a key intermediate result of this paper: the ``anthropic factor'' $h$ is irrelevant.  Its only possible effect is to further suppress the probability distribution in a regime whose integrated probability is already negligible, namely for $t_{\rm c} \ll t_{\rm obs}$ or $t_\Lambda \ll t_{\rm obs}$.  Near the half-lines of maximal probability, which dominate the distribution, $h$ will at most contribute factors of order unity, which we neglect here in any case, and which will not lift the degeneracy along the two half-lines, because they can only depend on the variable orthogonal to each line.

The other prefactor, $g(\log t_{\rm c})$, encodes the prior probability distribution over the time of curvature domination. We will assume that $g$ decreases mildly, like an inverse power of $\log t_{\rm c}$. (If slow-roll inflation is the dominant mechanism responsible for the delay of curvature domination, $\log t_{\rm c}$ corresponds to the number of $e$-foldings. If $g$ decreased more strongly, like an inverse power of $t_{\rm c}$, then inflationary models would be too rare in the landscape to explain the observed flatness.)  This will tilt the arrows slightly left in regions IV and V, lifting the degeneracy along the horizontal half-line. Because of the logarithmic dependence, the peak at $t_{\rm c} \sim t_{\rm obs}$ is very wide, so $t_{\rm c}$ could be large enough for curvature to be unobservable in vacua with $t_\Lambda \sim t_{\rm obs}$.  For example, with $g\sim (\log t_{\rm c}) ^{-4}$, the probability of detecting curvature in a future experiment with sensitivity $|\Omega_k|\sim 10^{-4}$ is only about 10\%~\cite{FreKle05}.

The degeneracy along the vertical half-line $t_{\rm c} \sim t_{\rm obs}$ would render the probability distribution unintegrable, {\em unless the effective number of vacua in the landscape, $\bar {\cal N}$, is finite}.   By ``effective'', we mean that vacua without observers, and vacua that are very suppressed cosmologically, are not included in $\bar{\cal N}$.  The effective spectrum of $\Lambda$ is discrete, with spacing of order $\bar {\cal N}^{-1}$, and the smallest positive value of $\Lambda$ will be of this order.  Correspondingly, there will be a largest value of $t_\Lambda$, $t_\Lambda^{\rm max}\sim \bar {\cal N}^{1/2}$.  This ``discretuum cutoff'' reduces the half-line of maximal probability to the interval
\begin{equation}
\log t_{\rm c} = \log t_{\rm obs}~,~~\log t_{\rm obs} \lesssim \log t_\Lambda \lesssim\log t_\Lambda^{\rm max} ~.
\label{eq-interval}
\end{equation}  

So far, we have treated the timescale of observers, $t_{\rm obs}$, as a fixed input parameter.  We will now extend our analysis to determine the probability distribution over all three parameters, including $\log t_{\rm obs}$.  Part of this work has already been done, since our calculation of the mass $M_{\rm CP}$ within the causal patch took into account the dependence on $t_{\rm obs}$ as well as on $t_\Lambda$ and $t_{\rm c}$.    The number of observers per unit mass, $h$, also includes a $t_{\rm obs}$-dependence, though we have yet to make any assumptions about it.  For the joint prior probability, we can write
\begin{equation}
\frac{d^3\tilde p}{d \log t_\Lambda d \log t_{\rm c} d \log t_{\rm obs}}=
f\,\frac{d\tilde p}{d\log t_\Lambda d\log t_{\rm c}} ~,
\label{eq-bay}
\end{equation}
where the last factor is given by Eq.~(\ref{eq-prior}), and $f\equiv p(t_{\rm obs}|t_\Lambda, t_{\rm c})$ is the conditional probability that observers exist at the time $t_{\rm obs}$ in a vacuum with parameters $(t_\Lambda, t_{\rm c})$.  We thus find a trivariate probability distribution simply by multiplying Eq.~(\ref{eq-pcppos})  by the function $f$.

To analyse this distribution, we will first marginalize over $\log t_\Lambda$ and $\log t_{\rm c}$ to determine the most probable value of $\log t_{\rm obs}$.  Evaluating Eq.~(\ref{eq-pcppos}) at this value then predicts $\log t_\Lambda$ and $\log t_{\rm c}$.   We may restrict the integral over $\log t_{\rm c}$ and $\log t_\Lambda$ to the neighborhood of the line interval (\ref{eq-interval}), which contains nearly all of the probability at any fixed $t_{\rm obs}$, and in which we argued that neither $f$ nor $h$ can depend significantly on $\log t_{\rm c}$ or $\log t_\Lambda$.   With $g\sim (\log t_{\rm c} )^{-k}$, $k>1$, we find from Eq.~(\ref{eq-pcppos}) that the marginal probability distribution over $\log t_{\rm obs}$ is $fh\,l\, t_{\rm obs} ^{-1}$, where $l$ is a rational function of $\log t_{\rm obs}$.  Let us assume that the product $fh$ grows more strongly than linearly with $t_{\rm obs}$, i.e., that $fh\sim t_{\rm obs}^{1+\epsilon}$,with $\epsilon>0$ ~\footnote{We know of no argument against this assumption.  Indeed, it seems clear that $h$ and $f$ should both increase with $t_{\rm obs}$, when averaged over many vacua with $t_{\rm obs} \sim t_{\rm c} \lesssim t_\Lambda$.  The density of matter at the time $t_{\rm obs}$ is of order $t_{\rm obs}^{-2}$, which implies that the maximum number of nonoverlapping quanta per unit mass grows as $t_{\rm obs}^{1/2}$~\cite{BouHar10}.  Supposing that a system of sufficient complexity to function as an observer requires a fixed minimum amount of quanta, this implies that more observers can be produced per unit mass, on average.  Similarly, the fraction of vacua containing observers should increase with $t_{\rm obs}$, since the number of elementary interactions that can take place in the universe, and hence, the probability that successful evolution of observers can take place, increases with time.  (This argument is due to Roni Harnik.)} With this assumption, the marginal distribution becomes $t_{\rm obs}^\epsilon l(\log t_{\rm obs})$. This favors large values for $t_{\rm obs}$: $\log t_{\rm obs}\sim \log t_\Lambda^{\rm max}$.  At this value of $\log t_{\rm obs}$, the interval of maximum probability (\ref{eq-interval}) for $(\log t_\Lambda,\log t_{\rm c})$ shrinks to a single point, and we predict
\begin{equation}
\log t_\Lambda \approx \log t_{\rm c} \approx \log t_{\rm obs} \approx \log(\bar {\cal N}^{1/2})~.
\label{eq-final}
\end{equation}

We thus predict that the number of vacua that contain observers and can be cosmologically produced is $\bar {\cal N} \sim 10^{123}$.  The relation between $\bar {\cal N}$ and the total number of vacua in the landscape is not trivial.  String theory contains infinitely many stable supersymmetric vacua, for example vacua of the form $AdS_p\times S^q$, with arbitrary integer flux.  Since we do not live in such a vacuum, it must be the case that they either cannot contain any observers, or that their integrated cosmological production rate is finite, for example because the production of very large flux values is highly suppressed.  If the speculation that ten-to-the-millions of vacua arise from F-theory constructions~\cite{DenDou06} holds up, then similar considerations would have to apply to explain why $\log \bar {\cal N}$ is so much smaller.

This work was supported by the Berkeley
Center for Theoretical Physics, by the National Science Foundation
(award number 0855653), by fqxi grant RFP2-08-06, and by the US
Department of Energy under Contract DE-AC02-05CH11231. VR is supported
by an NSF graduate fellowship.

\bibliography{all}

\end{document}